# Dirac Nodal Lines and Nodal Loops in a Topological Kagome Superconductor CsV$_3$Sb$_5$

Zhanyang Hao[1,2,*], Yongqing Cai[1,2,*], Yixuan Liu[1,2,*], Yuan Wang[1,2,*], Xuelei Sui[3], Xiao-Ming Ma[1,2], Zecheng Shen[1,2], Zhicheng Jiang[4], Yichen Yang[4], Wanling Liu[4], Qi Jiang[4], Zhengtai Liu[4], Mao Ye[4], Dawei Shen[4], Yi Liu[5], Shengtao Cui[5], Jiabin Chen[3], Le Wang[1,2], Cai Liu[1,2], Junhao Lin[1], Jianfeng Wang[6,3,#], Bing Huang[3], Jia-Wei Mei[1,2,7,#] and Chaoyu Chen[1,2,#]

[1] Shenzhen Institute for Quantum Science and Engineering, and Department of Physics, Southern University of Science and Technology, Shenzhen 518055, China

[2] International Quantum Academy, and Shenzhen Branch, Hefei National Laboratory, Futian District, Shenzhen 518048, China

[3] Beijing Computational Science Research Center, Beijing 100193, China

[4] State Key Laboratory of Functional Materials for Informatics and Center for Excellence in Superconducting Electronics, Shanghai Institute of Microsystem and Information Technology, Chinese Academy of Sciences, Shanghai 200050, China

[5] National Synchrotron Radiation Laboratory, University of Science and Technology of China, Hefei 230029, China

[6] School of Physics, Beihang University, Beijing 100191, China

[7] Shenzhen Key Laboratory of Advanced Quantum Functional Materials and Devices, Southern University of Science and Technology, Shenzhen 518055, China

[*] These authors contributed equally to this work.

[#] Correspondence should be addressed to J.W. (wangjf06@buaa.edu.cn), J.M. (meijw@sustech.edu.cn) and C.C. (chency@sustech.edu.cn)

## Abstract

The intertwining of charge order, superconductivity and band topology has promoted the $A$V$_3$Sb$_5$ ($A$=K, Rb, Cs) family of materials to the center of attention in condensed matter physics. Underlying those mysterious macroscopic properties such as giant anomalous Hall conductivity (AHC) and chiral charge density wave is their nontrivial band topology. While there have been numerous experimental and theoretical works investigating the nontrivial band structure and especially the van Hove singularities, the exact topological phase of this family remains to be clarified. In this work, we identify CsV$_3$Sb$_5$ as a Dirac nodal line semimetal based on the observation of multiple Dirac nodal lines and loops close to the Fermi level. Combining photoemission spectroscopy and density functional theory, we identify two groups of Dirac nodal lines along $k_z$ direction and one group of Dirac nodal loops in the $A-H-L$ plane. These nodal loops are located at the Fermi level within the instrumental resolution limit. Importantly, our first-principle analyses indicate that these nodal loops may be a crucial source of the mysterious giant AHC observed. Our results not only provide a clear picture to categorize the band structure topology of this family of materials, but also suggest the dominant role of topological nodal loops in shaping their transport behavior.

**Introduction:**

At the heart of the current condensed matter physic lies the interplay between band topology and certain symmetry breaking orders [1], as such interplay could give rise to novel quantum phases and phenomena such as quantum anomalous Hall effect in Chern insulator [2-5], Majorana mode in topological superconductor [6-10], axion dynamics in axion insulator [11,12] and so on. In this context, the recently discovered $A$V$_3$Sb$_5$ ($A$=K, Rb, Cs) family of materials [13] have attracted tremendous research interests due to their coexisting nontrivial band topology arising from kagome lattice geometry, unconventional superconductivity, and charge density wave (CDW). These materials share the same layered structure with two-dimensional (2D) transition metal V kagome lattice [13-15], resulting in multiple Dirac cones and the rich nature of van Hove singularities in their electronic structure as demonstrated via angle-resolved photoemission spectroscopy (ARPES) [16-29] measurements and density-functional theory (DFT) calculations [30]. Superconductivity was discovered with $T_C = 0.93\ K$ for KV$_3$Sb$_5$ [14], $T_C = 0.92\ K$ for RbV$_3$Sb$_5$ [15], and $T_C = 2.5\ K$ for CsV$_3$Sb$_5$ [16], respectively. Meanwhile, at $T^* \sim 80 - 100\ K$ a CDW order was observed [13-16,18,31,32], whose relationship with superconductivity is still under debate [20,31,33-38]. Nevertheless, recent experimental and theoretical works have suggested the unconventional nature of both orders [18,31,37-41].

Surprisingly, giant anomalous Hall conductivity (AHC) up to the order of $10^4\ \Omega^{-1}cm^{-1}$ was observed in nonmagnetic KV$_3$Sb$_5$ [42] and CsV$_3$Sb$_5$ [39]. Recent muon spin spectroscopy[43] measurement demonstrates the absence of local V$^{4+}$ moments. So the AHC has no magnetic origin, but may be associated with the chiral CDW order [31,39,42,44]. In this sense, it is of key importance to correctly understand their topological band structure for the electronic origin of the giant AHC. It was first proposed that this family of materials can be categorized as $\mathbb{Z}_2$ topological metals with a continuous direct gap [16]. Subsequent ARPES measurements have revealed multiple Dirac crossings and the rich nature of van Hove singularities (VHSs) close to the Fermi level ($E_F$) [16-28]. Nevertheless, under the combined $\mathcal{PT}$ symmetry with inversion symmetry $\mathcal{P}$ and time reversal symmetry $\mathcal{T}$, this system should contain Dirac nodal lines rather than isolated Dirac points [30] in its three-dimensional (3D) Brillouin zone (BZ). This suggests that the exact topological phase of $A$V$_3$Sb$_5$ ($A$=K, Rb, Cs) family remains to be clarified and the necessity of a full 3D BZ band investigation to identify its topological band features.

In this work, we present a comprehensive ARPES spectral mapping covering the full 3D BZ of topological kagome superconductor CsV$_3$Sb$_5$. Combining ARPES and DFT calculation, we identify two types of Dirac nodal lines in a shallow energy region ($\leq 0.3\ eV$) below $E_F$. The first type consists of two groups of nodal lines along $k_z$ direction, arising from the quasi-2D nature of the kagome lattice. The second type consists of one group of nodal loops surrounding the $H$ points in $k_x - k_y$ plane, i.e., $A - H - L$ plane. The Dirac energies of these nodal loops are almost equal to $E_F$ within the instrumental resolution limit. The satisfactory agreement between ARPES and DFT fully corroborates the existence of these low-energy Dirac nodal lines and loops. Importantly, our first-principle analyses indicate that these nodal loops may provide dominant source of Berry curvature and thus the giant AHC. Our results not only represent a clear clarification of the topological band structure in the kagome $A$V$_3$Sb$_5$ ($A$=K, Rb, Cs) family of materials, but also suggest nodal loops as the electronic source of the exotic transport response of the host materials.

**Results:**

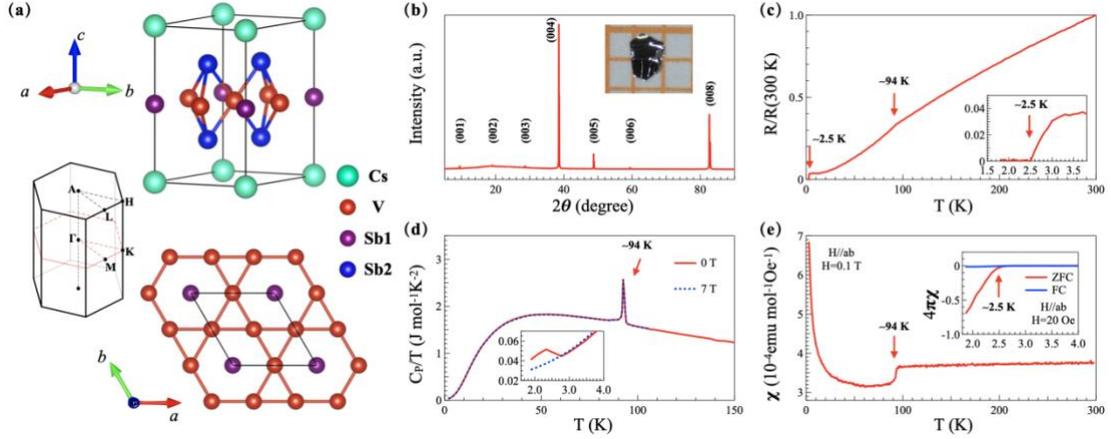

**Fig. 1. Crystal structure and characterization of single-crystalline CsV$_3$Sb$_5$.** (a) Crystal structure of CsV$_3$Sb$_5$. The V atoms construct a kagome lattice, the Sb1 atoms are in the center of the V hexagon and the Sb2 atoms form a honeycomb lattice above and below the kagome layer. The left inset shows the corresponding 3D BZ and the high-symmetry points. (b) Single-crystal XRD pattern of the (00$l$) plane. The inset shows the optical image of the CsV$_3$Sb$_5$ crystal placed on the 1 mm grid paper. (c), (d) and (e) Temperature-dependent resistivity ratio, heat capacity, and magnetic susceptibility measurements of CsV$_3$Sb$_5$ single crystal. Phase transitions indicating the charge density wave can be observed at around 94 $K$. All the three insets exhibit the bulk superconductivity transition around 2.5 $K$.

Single crystals of CsV$_3$Sb$_5$ adapt a kagome structure of V atoms with a space group of *P6/mmm* [13], as shown in Fig. 1(a). The Sb1 atoms are in the center of the V hexagons and the Sb2 atoms compose the antimony layers which are sandwiched by the V kagome layers and Cs layers. The X-ray diffraction (XRD) pattern of (00$l$) plane is exhibited in Fig. 1(b), which indicates the high crystalline quality. The lattice parameter of the *c*-axis obtained from XRD data is about 9.308 Å, which is also consistent with the previous study [13]. For ARPES measurement, the cleavage plane is parallel to the *ab* plane (inset of Fig. 1(b)) and perpendicular to the *c* axis direction.

In Fig. 1(c), an obvious kink at ~94 $K$ can be seen from the curve which is related to the CDW phase transition, and an apparent bulk superconducting transition feature exists at ~2.5 $K$ (inset of Fig. 1(c)). Similarly, detailed temperature-dependent heat capacity and magnetization measurements were performed systematically. From Fig. 1(d), one can see that a sharp CDW transition peak at ~94 $K$ exists for both 0 $T$ and 7 $T$ magnetic field. In contrast, the superconductivity transition occurs at ~2.5 $K$ under the 0 $T$ magnetic field and is suppressed under 7 $T$ magnetic field. For the magnetic susceptibility data in Fig. 1(e), under a 0.1 $T$ magnetic field, there is an obvious drop at ~94 $K$. The low temperature magnetic susceptibility with field-cooling (FC) and zero-field-cooling (ZFC) measurements at 20 $Oe$ clearly shows the onset of superconductivity at ~2.5 $K$ (inset of Fig. 1(e)). Note that the magnetic field is parallel to the *ab* plane in magnetization measurement and along the *c*-axis in heat capacity measurement. For ARPES measurement through this work, the sample temperature was kept at 10 $K$ unless specified.

It is noted that while the structural signature of CDW has been clearly observed [31,32,36,40,41,45-47], the corresponding electronic signature, *e.g.*, folded band, is absent in most of the ARPES measurements [16-27]. In this work, we shall treat the ARPES spectra in and out of the CDW phase equally.

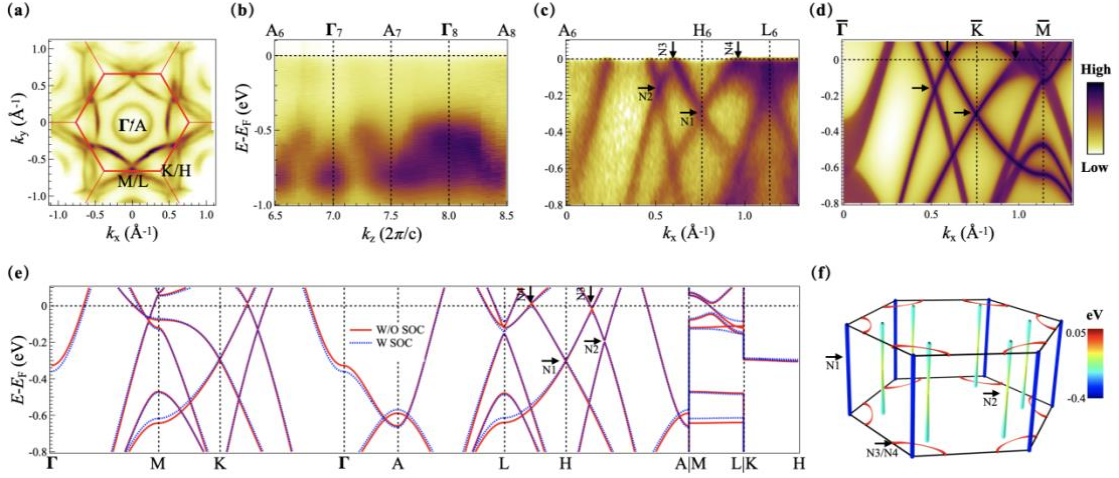

**Fig. 2. General electronic structure with multiple Dirac nodal lines and loops in CsV$_3$Sb$_5$ revealed by ARPES observations and DFT calculations.** (a) Fermi surface mapping in 2D BZ, where the red lines plot the BZ boundaries with high-symmetry points indicated. (b) Photon energy-dependent spectra for photoemission intensity from the BZ center $\Gamma/A$ points. (c) Spectra taken along momentum cuts $A_6 - H_6 - L_6$. Black arrows indicate the existence of 4 Dirac nodes labeled as "N1" to "N4". (d) Spectra along the $\overline{\Gamma} - \overline{K} - \overline{M}$ path from the projection of DFT calculated bands onto surface BZ. (e) DFT calculated electronic structure along all high-symmetry paths. Red solid lines represent bands without spin-orbit coupling (SOC) while blue dashed lines are results with SOC included. (f) Visualization of Dirac nodal lines and loops formed by nodes N1, N2, N3 and N4 in the 3D BZ from DFT calculations. The different colors represent their energy relative to $E_F$. Note that the calculated $E_F$ is slightly shifted upward in order to match the experimental results.

The layered nature of the kagome lattice in CsV$_3$Sb$_5$ makes its band dispersion along $k_z$ direction quite weak. While the DFT calculations predict a $\sim 350\ meV$ dispersion along $\Gamma - A$ direction [16,30,48], such $k_z$ dependence cannot be found from recently reported ARPES results [16-27]. It is thus of importance to examine this discrepancy. Judging from the Fermi surface geometry, we align the ARPES momentum cut to $\overline{\Gamma} - \overline{K} - \overline{M}$ high-symmetry path ($k_y = 0$ path in Fig. 2(a)) and perform photon energy dependent measurement for a photon energy range of $70 - 130\ eV$. As shown in Fig. 2(b), the spectral intensity from $\overline{\Gamma}$ point exhibits a clear periodic dispersion, which corresponds to the $\Gamma - A$ dispersion as predicted by DFT calculations in Fig. 2(e) [16,30,48]. This periodicity helps us to identify the bulk $\Gamma$ and $A$ points of the 3D BZ as specified in Fig. 2(b). In this way, the $k_z$ location of all the following ARPES spectra cuts and $k_x - k_y$ mappings can be approximately determined.

ARPES spectra along the high-symmetry cut $A_6 - H_6 - L_6$ is shown in Fig. 2(c). Projection of DFT calculated dispersion onto the surface BZ path $\overline{\Gamma} - \overline{K} - \overline{M}$ is also presented in Fig. 2(d) for comparison. Both ARPES and DFT results show consistent features as discussed below. One parabolic electron pocket can be found around $\Gamma/A$, whose band bottom contributes to the $k_z$ dispersion observed in Fig. 2(b). Besides that, all the bands are linearly dispersed in a narrow energy region below $E_F$. These bands cross each other and form Dirac nodes. We have identified as many as four Dirac nodes with Dirac energy $E_D \leq 0.3\ eV$, as indicated by black arrows in Figs. 2(c, d). Sorted according to binding energy, these nodes are labeled as "N1" up to "N4". Dirac node N1 represents the Dirac cone at $K/H$ point with $E_D \approx 0.3\ eV$ while N2 the Dirac cone between $\Gamma/A$ and $K/H$ points with $E_D \approx 0.15\ eV$. When moving to $E_F$, two nodes, N3 and N4, appear. Node N3 comes from the crossing between the upper Dirac cones of N1 and N2, and N4 shares a similar origin. The exact energy position of N3 and N4 may be slightly below or up $E_F$, yet we consider them effectively located at $E_F$ since their Dirac energy is comparable to or even smaller than our practical energy resolution limit and the energy distribution width of ARPES measured bands.

The emergence of multiple Dirac nodes as discussed above has been captured in recent ARPES results [16-27] but their $k_z$ dependence in 3D BZ lacks systematic discussion. In Fig. 2(e), we present the DFT calculated electronic structure along all the high-symmetry paths in the 3D BZ of CsV$_3$Sb$_5$. From the calculations, for a narrow energy range close to $E_F$, while the bands along $\Gamma - A$ and $M - L$ possess a clear $k_z$ dispersion, the bands along $K - H$ are almost $k_z$ independent. This brings about two important electronic features. On the one hand, the Dirac nodes N1 and N2 look almost the same at paths $\Gamma - K - M$ and $A - H - L$, suggesting the formation of Dirac nodal lines along $k_z$ direction in the 3D BZ. On the other hand, the nodes N3 and N4 are located not only at almost the same energy, but also at the same upper Dirac cone of node N1, suggesting the possibility of one closed nodal loop at the Fermi level surrounding $H$ point. In Fig. 2(f), the DFT predicted distributions of these four Dirac nodes are shown in the 3D BZ. It is clearly demonstrated that there exist two groups of straight nodal lines along the $k_z$ direction and one group of closed nodal loops surrounding $H$ point at the $A - H - L$ plane.

From a point of view of symmetry, these three groups of nodal lines/loops are protected by different symmetries. The CsV$_3$Sb$_5$ crystal has a point group of $D_{6h}$, including inversion $I$, $C_6$ rotation, six vertical mirrors $\sigma_v$ and one horizontal mirror $\sigma_h$ symmetries. The nodal lines formed by N1 (blue in Fig. 2(f)) are fixed along $K - H$, which are protected by the $C_{3v}$ subgroup along this high-symmetry line; while the nodal lines formed by N2 (nearly green in Fig. 2(f)) are in the three $\Gamma - K - H - A$ planes and protected by the $\sigma_v$ mirror planes. The nodal loops formed by N3/N4 surrounding $H$ point at $A - H - L$ plane (red in Fig. 2(f)) are protected by the $\sigma_h$ mirror symmetry. Any local perturbations preserving the related symmetries will not break these nodal lines/loops. It is noted that the SOC effect is weak, and it only opens a small gap for the Dirac nodes and nodal lines/loops (see Fig. 2(e)), consistent with the previous DFT works [16,30,48]. In the following, we will use ARPES data to visualize these nodal lines and nodal loops separately.

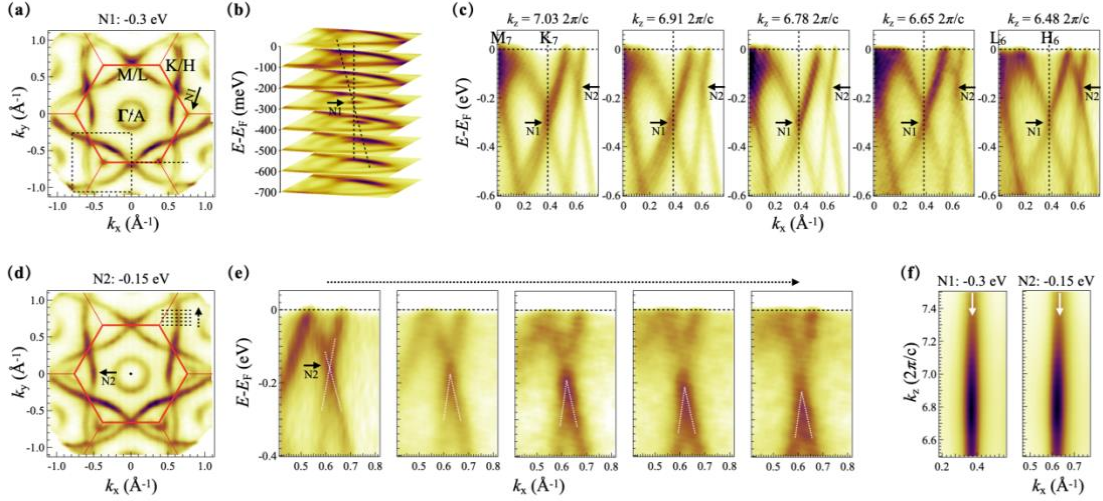

**Fig. 3. Visualizations of Dirac nodal lines along $k_z$ direction formed by nodes N1 and N2.** (a) Constant energy contours (CECs) in $k_x - k_y$ plane at a binding energy of $\sim 0.3\ eV$, which shows the Dirac nodal point (black arrow) formed by node N1. (b) CECs in $k_x - k_y$ plane at a series of binding energies. Black dashed lines indicate the N1 Dirac cone dispersion. (c) Spectra taken along a high-symmetric $M/L - K/H - \Gamma/A$ direction at a series of $k_z$ planes with black arrows highlighting the positions of N1 and N2. (d) CEC in $k_x - k_y$ plane at a binding energy of $\sim 0.15\ eV$, which shows the Dirac nodal point (black arrow) formed by node N2. (e) Spectra from a series of momentum cuts whose positions are indicated by black dashed lines in (d). White dashed lines highlight the gap opening of the Dirac cone when moving away from node N2. (f) Direct visualizations of N1, N2 Dirac nodal lines in the $k_z - k_x$ plane from the CECs at the corresponding binding energies.

Concerning the nodal lines along $k_z$ direction, we start with node N1. Fig. 3(a) presents the constant energy contours (CECs) at its binding energy of $\sim 0.3\ eV$, which shows a nodal point at $K/H$ point as indicated by the black arrow. By analyzing its shape evolution in CECs at a series of binding energies (Fig. 3(b)), we find that the node N1 forms a triangularly deformed Dirac cone in $k_x - k_y$ plane. The evolution of this Dirac cone along $k_z$ direction is presented in Fig. 3(c) at five selected $k_z$ planes. Despite some intensity variations, the dispersion of N1 Dirac cone shows no noticeable changes from $K_7$ to $H_6$, suggesting the existence of N1 nodal line along $k_z$ direction. In the left panel of Fig. 3(f), the CEC at its corresponding energy of $\sim 0.3\ eV$ in $k_x - k_z$ plane directly visualizes this N1 nodal line.

For node N2, its CEC in $k_x - k_y$ plane shows a deformed, closed hexagon in the first 2D BZ (Fig. 3(d)), suggesting a nodal loop formation in $k_x - k_y$ plane. However, by examining its detailed dispersion along different momentum cuts as shown in Fig. 3(e), we find that the gapless N2 Dirac cone only exists along the high-symmetric $\overline{\Gamma} - \overline{K}$ path. When moving away from this high-symmetry path, this Dirac cone gradually opens a gap. This suggests that the node N2, like N1, also forms nodal points in $k_x - k_z$ plane. The $k_z$ evolution of N2 shown together with N1 in Fig. 3(c) draws the same conclusion that the nodal lines are formed along $k_z$ direction. In the right panel of Fig. 3(f), the CEC at its corresponding energy of $\sim 0.15\ eV$ in $k_x - k_z$ plane directly visualizes this N2 nodal line.

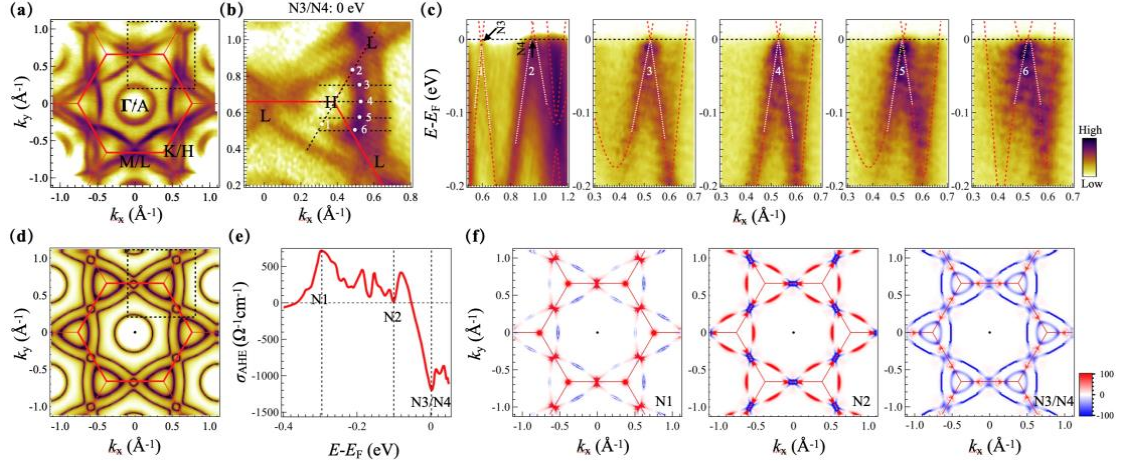

**Fig. 4. Visualizations of Dirac nodal loops at Fermi surface formed by nodes N3/N4 and calculated AHC and Berry curvatures contributed by the observed nodal lines/loops.** (a) CEC in $k_x - k_y$ plane at $E_F$. (b) CEC in the momentum region indicated by black dashed box in (a). White dots labeled by numbers indicate the nodes determined from spectra analysis in (c). (c) Spectra taken from a series of momentum cuts whose positions are indicated by black dashed lines in (b). White dotted lines with labeled numbers indicate the dispersion evolution and momentum evolution of node N3/N4. DFT calculated dispersion evolution along the same momentum cuts are demonstrated by the red dashed lines. (d) Simulated Fermi surface by DFT calculations. (e) Calculated AHC as a function of binding energy. (f) Calculated Berry curvature distributions in $A - H - L$ plane at different binding energies of $-0.3\ eV$, $-0.1\ eV$ and $0\ eV$, respectively.

Concerning Dirac nodes N3 and N4, we first analyze their CEC at the Fermi level in the $A - H - L$ plane. Note that from the DFT calculated dispersion (Fig. 2(e)), the node N4 doesn't exist in the $\Gamma - K - M$ plane. Fig. 4(b) highlights the Fermi surface in the momentum region surrounding the $H$ point in the CEC, as indicated by black dashed box in Fig. 4(a). In this CEC, one can see two closed triangles surrounding the $H$ point. While the outer triangle comes from the upper Dirac band of node N2, the inner one is formed by the Dirac node N3/N4. To fully demonstrate this, we extract several spectra from a series of momentum cuts as denoted by the black dashed lines in Fig. 4(b). As shown in Fig. 4(c), the linear dispersions are observed and highlighted by the white dotted lines in these spectra. All show the signatures of Dirac nodes close to the Fermi level, which are labeled by numbers in white. These Dirac nodes are connected together to form a closed loop (Fig. 4(b)) considering the three-fold rotation symmetry. In Fig. 4(d), we plot the simulated Fermi surface by the DFT calculations in the $A - H - L$ plane. The satisfactory agreement between ARPES (Fig. 4(a)) and DFT (Fig. 4(d)) fully corroborates the existence of Dirac nodal loops at the Fermi surface.

Finally, we discuss the possible contributions of these Dirac nodal loops/lines to the exotic giant AHC observed by recent experiments[39,42]. Although there is no observation of magnetic order, the unconventional chiral CDW of CsV$_3$Sb$_5$ can break the time-reversal symmetry[32,43], leading to the appearance of anomalous Hall effect. A modal estimation based on the gapped Dirac nodes around $M$ point gives rise to AHC of $10^2\ \Omega^{-1} cm^{-1}$[31], which is two orders of magnitude lower than the observed value. Here, we emphasize that the nodal loops/lines observed here can contribute more additional Berry curvatures and greater AHC once the time-reversal symmetry is broken, similar to Co$_3$Sn$_2$S$_2$ [49]. To demonstrate this point, a ferromagnetic state of CsV$_3$Sb$_5$ is employed to simulate the situation of time-reversal symmetry breaking, which can be achieved by DFT+$U$ calculations with a Hubbard $U = 2\ eV$ (the average local moment on each V atom is $0.27\ \mu_B$, see method section). The calculated AHC under different binding energies is shown in Fig. 4(e). Large AHC is achieved at some binding energies, especially near $E_F$ which can reach to a giant value of $10^3\ \Omega^{-1}cm^{-1}$. Here, three typical binding energies ($-0.3, -0.1$ and $0\ eV$) are selected to visualize the Berry curvature distributions in the BZ, as shown in Fig. 4(f). At the binding energy of $-0.3\ eV$, the calculated AHC is $\sim 750\ \Omega^{-1}cm^{-1}$, and the Berry curvatures are mainly distributed around the $L$ ($M$) and $H$ ($K$) points, reflecting the contributions of N1 nodal lines. At the binding energy of $-0.1\ eV$, the AHC is calculated nearly zero, arising from the cancellation of the Berry curvatures from the positive contributions of N2 nodal lines and negative contributions around the $L$ ($M$) point. At the Fermi level, the calculated AHC is as high as $\sim -1250\ \Omega^{-1}cm^{-1}$, and the nodal loops formed by N3/N4 play a dominant role in the Berry curvature distributions. Therefore, these Dirac nodal loops contribute to the large Berry curvature, which is directly responsible for the giant AHC observed in $A$V$_3$Sb$_5$ family of materials.

**Discussion:**

While numerous ARPES and DFT works have focused on the detailed band structure of $A$V$_3$Sb$_5$ family of materials, particularly the VHSs at the $\overline{M}$ point as they are closely related to the unconventional CDW order, it is still necessary to have a clear picture concerning their global electronic structure and the exact topological phase. By examining the electronic structure of CsV$_3$Sb$_5$ in its full 3D BZ from both experimental and theoretical aspects, we could identify it as a Dirac nodal line semimetal, especially with one group of nodal loops at the Fermi surface. Our results not only provide a clear picture to categorize the band structure topology of this family of materials, but also suggest the potential role of Dirac nodal loops in shaping their transport behavior . Concerning the giant AHC observed in this family [39,42], the lack of local or global magnetic moments suggests the intrinsic nature of AHC and the key role of electronic structure[13,43]. Under the chiral CDW order, besides the Dirac nodes around the $M$ point which gives rise to AHC of two orders of magnitude lower than the observed value [31], our results point out that the presence of multiple nodal loops at the Fermi level around bulk $H$ point could contribute additional larger Berry curvature and bring forth AHC whose value can reach as high as $\sim 1250\ \Omega^{-1}cm^{-1}$. This offers new insight into the intrinsic origin of giant AHC in this family and calls for further theoretical and experimental investigations.

**ACKNOWLEDGEMENTS**

We thank Prof. Haizhou Lu, Qihang Liu, Wen Huang for helpful discussions. This work is supported by National Natural Science Foundation of China (NSFC) (Grants Nos. 12074163 and 12004030), the Shenzhen High-level Special Fund (Grants No. G02206304 and No. G02206404), the Guangdong Innovative and Entrepreneurial Research Team Program (Grants No. 2017ZT07C062 and No. 2019ZT08C044), Shenzhen Science and Technology Program (Grant No. KQTD20190929173815000), the University Innovative Team in Guangdong Province (No.

**Materials and Methods**

**Sample growth and characterization**

$CsV_3Sb_5$ single crystals were synthesized by the self-flux method. High purity Cs (clump), V (powder) and Sb (ball) were mixed with a ratio of 2:1:6 in the glovebox and placed into an alumina crucible. The crucible was then double sealed into an evacuated quartz tube under high pressure to avoid oxidization during the reaction. In the furnace, the double sealed quartz tube was heated up to 500°C and kept at this temperature for 10 hours. Then the temperature was increased to 1050°C, after 12 hours the temperature was slowly decreased down to 650°C in 200 hours. After the heating procedure, the quartz tube was taken into the centrifuge to remove the excess flux. Many millimeter-sized hexagonal $CsV_3Sb_5$ single crystals were obtained.

The structure of the crystals was determined by x-ray diffraction with Cu Kα radiation at room temperature using a Rigaku MiniFlex diffractometer. The diffraction pattern can be well indexed by the (00l) reflections.

**Transport, heat capacity and magnetic measurements**

Magnetization measurements were performed by Magnetic Property Measurement System (MPMS3, Quantum Design). Quartz paddle was used, and the magnetic field was perpendicular to the $c$ - axis to minimize the diamagnetic contribution. Heat capacity and resistivity measurements were performed by Physical Property Measurement System (PPMS, Quantum Design). For resistivity, a standard four-probe method was employed with a probe current of 0.5mA. And the electrical current was in the $ab$ plane. For heat capacity, the selected single crystal was approximately 1.5 mm width and length, and Apezion N-grease was used to ensure the connection with heat capacity stage.

**ARPES measurement**

ARPES measurements were performed at the BL03U beamline of the Shanghai Synchrotron Radiation Facility and the BL-13U beamline of the National Synchrotron Radiation Laboratory (NSRL) both with DA30L electron analysers. The energy and angular resolution were set at 20 meV and less than 0.05°, respectively. Samples were cleaved *in situ* under ultra-high vacuum conditions with pressure better than $5 \times 10^{-11}$ mbar and temperatures below 20 K.

**First-principles calculations**

The first-principles calculations are performed using the Vienna ab initio simulation package [50] within the projector augmented wave method [51] and the generalized gradient approximation of the Perdew-Burke-Ernzerhof [52] exchange-correlation functional. The plane-wave basis with an energy cutoff of 400 eV and the $\Gamma$-centered 12×12×6 $k$-point meshes are adopted. Employing the experimental lattice constants of $a=b=5.495$ Å and $c=9.308$ Å, the crystal structure of $CsV_3Sb_5$ is relaxed with van der Waals correction [53] until the residual forces on each atom is less than 0.005 eV/Å. The ferromagnetic and 120°-antiferromagnetic states with different Hubbard $U$ are tested, which gives a ferromagnetic ground state with a small spin polarization energy when $U>0.8$ eV. The SOC effect is also considered in part of our calculations. A tight-binding (TB) Hamiltonian based on the maximally localized Wannier functions (MLWF) [54] is constructed to get the energy eigenvalues and eigenstates for further Fermi surface plot using the WannierTools package [55]. In order to simulate the time-reversal symmetry breaking to calculate the AHC of CsV$_3$Sb$_5$, a ferromagnetic ground state is employed by adopting DFT+$U$ calculations with $U = 2\ eV$, which gives a spin polarization energy of $0.1\ eV/unit\ cell$ and an average local moment of $0.27\ \mu_B/V$. Then, the AHC and Berry curvatures are calculated by constructing MLWF and using WannierTools package.